\documentclass[12pt]{iopart}

\usepackage{graphicx}
\usepackage[colorlinks]{hyperref}
\hypersetup{
    linkcolor=black,       
    citecolor=black,       
    filecolor=black,       
    urlcolor=black         
}

\usepackage{harvard}
\usepackage{iopams}  

\begin{document}

\title[Proton-Acoustic Wave Effects on Proton Transverse Heating]{Proton-Acoustic Wave Effects on the Relaxation of Proton Transverse Heating in Magnetized Plasmas}

\author{Martín A. Quijada$^{1}$, Pablo S. Moya$^{2}$ and Roberto E. Navarro$^{1}$}

\address{%
$^{1}$Departamento de Física, Facultad de Ciencias Físicas y Matemáticas, Universidad de Concepción 4070386, Chile\\
$^{2}$Departamento de Física, Facultad de Ciencias, Universidad de Chile, Las Palmeras 3425, Ñuñoa, Santiago 7800003, Chile}
\ead{mquijada2018@udec.cl, pablo.moya@uchile.cl, roberto.navarro@udec.cl}
\vspace{10pt}
\begin{indented}
\item[]\today
\end{indented}

\begin{abstract}
  Transverse electromagnetic and electrostatic plasma wave modes propagating along a background magnetic field $\vec{B}_0$ are
  independent according to linear kinetic theory. However, resonant
  interactions and energy exchange between waves and particles break
  this linear decoupling. This work tracks the coupled evolution of
  Alfvén-cyclotron (ACWs) and Ion-acoustic waves (IAWs) by solving
  moment-based quasilinear equations for a collisionless plasma of
  bi-Maxwellian protons and Maxwellian electrons.  Unlike earlier
  quasilinear studies that adopt the cold-electron limit, our
  formulation retains the full kinetic response of both species,
  treating the electrons as a thermal reservoir to isolate proton
  heating.  A parameter survey over
  $0.01\leq\beta_{\parallel p}\leq10$ and $1\le T_e/T_p\le10$ shows
  that an ambient spectrum of ACWs can drive significant perpendicular proton heating and raise the temperature anisotropy from initially isotropic conditions at low $\beta_{\parallel p}\lesssim0.1$, thereby triggering cyclotron instabilities. The quasilinear evolution self-regulates the ACW, driving the system toward a quasi-stationary state with $\gamma/\Omega_p<10^{-1}$ and reduced anisotropy. As $T_e/T_p$ increases, IAWs become less damped and absorb a larger share of the fluctuation energy through Landau resonance, reducing the efficiency of ACW-driven proton heating and thus regulating the instability. For sufficiently large $\beta_{\parallel p}$ or $T_e/T_p\gtrsim5$, ACWs become inefficient drivers of perpendicular heating, leaving IAWs as the dominant dissipation channel. These results explain how modest electrostatic activity in low-$\beta$ environments such as the inner heliosphere and planetary magnetosheaths can regulate, but not indefinitely sustain, cyclotron instabilities. While the moment-based quasilinear approach neglects phase-space diffusion and BGK trapping, the framework cleanly separates the roles of IAWs and ACWs and can be extended to kinetic electron sub-populations or other instabilities.
\end{abstract}

%
\vspace{2pc}

\noindent{\it Corresponding Author}: Roberto E. Navarro (roberto.navarro@udec.cl).

\submitto{\PS}
%
%
%

\section{Introduction}

Moment-based quasilinear (MbQL) theory provides a tractable framework
for following the slow relaxation of temperature-anisotropy-driven
electromagnetic ion-cyclotron (EMIC)
instabilities. \citeasnoun{davidson1975} provided one of the earliest
dedicated moment-based studies for purely transverse Alfvén waves
propagating in proton-electron plasmas. They showed that the proton
temperature anisotropy $T_\perp/T_\parallel$, with $T_\perp$ and
$T_\parallel$, the perpendicular and parallel proton temperatures with
respect to a background magnetic field $\vec{B}_0$, is ultimately
regulated by the EMIC instability, and energy isotropisation proceeds
on a timescale of only a few inverse linear growth rates
$\gamma^{-1}$. Building on the same idea, \citeasnoun{seough2012}
studied the asymptotic MbQL state of both EMIC and firehose
instabilities, finding that these instabilities eventually relax to a
state close to either of the instability thresholds, described by an
inverse relationship between the proton temperature anisotropy
$T_\perp/T_\parallel$ and the plasma beta
$\beta_\parallel = 8\pi n k_B T_\parallel/B_0^2$, where $n$ is the
plasma density and $k_B$ the Boltzmann constant. They also found that
magnetic fluctuations develop with increasing amplitudes as
$\beta_\parallel$ increases, which is consistent with
observational~\cite{Bale2009} and
theoretical~\cite{camporeale2010,Navarro2014} studies of the solar
wind. 

However, EMIC instabilities can coexist with other instability sources. For example, mirror and EMIC instabilities have been observed
simultaneously in the dawnside and duskside of the Earth's outer
magnetosphere~\cite{liu2024}, Saturn magnetosheath~\cite{duanmu2023},
and predicted in numerical simulations of fast magnetosonic shocks in
the solar wind~\cite{lee2017} and the Earth's
magnetosheath~\cite{Hellinger2003,shoji2012}. Under their combined
effects, MbQL theory predicts time-asymptotic plasma states where both
instabilities are relaxed, settling near the mirror instability
threshold for high beta, but to the EMIC threshold for low
beta~\cite{yoon_quasilinear_2012,yoon_proton_2021}. On the other hand,
\citeasnoun{yoon2017} considered the MbQL dynamical interplay between
protons and electrons, finding that parallel electron firehose
counter-balance EMIC instabilities, preventing the uniform progression
of solar wind protons toward a marginal state. \citeasnoun{moya_toward_2021} also considered the MbQL temporal
evolution of electron kappa distributions, showing that the reduction of
plasma instabilities can be accompanied by an increment of
high-energy tails.

The presence of instabilities alone cannot explain why most of the
solar wind~\cite{kasper2002,hellinger_solar_2006,Bale2009},
magnetosphere~\cite{espinoza2022}, or even certain space-inspired laboratory plasmas~\cite{Keiter2000,beatty2020} are commonly found in states far from marginal stability and close to temperature isotropy.
To address this, \citeasnoun{seough2013} introduced a time-varying ambient field into the MbQL and found that initially unstable states can relax to subthreshold conditions bounded by mirror and oblique firehose instabilities. \citeasnoun{moya_effects_2021} explicitly included a
background spectrum of Alfvénic turbulence in the MbQL initial conditions, mimicking solar-wind fluctuations, and found that sufficient ambient wave power can drive an initially stable or isotropic proton distribution into an unstable and heated state. On the other hand, \citeasnoun{seough2023} formulated the first self-consistent MbQL expanding-box model for the solar wind and found that evolutionary paths of $\beta_\parallel$ and $T_\perp/T_\parallel$ closely overlay the proton $T_\perp/T_\parallel$ evolution from 0.3 to 2.5 AU~\cite{marsch_temperature_2004,matteini_evolution_2007}.

Most MbQL studies remain restricted to the effects of transverse
electromagnetic fluctuations. However, spacecrafts routinely observe
broadband electric-field signatures attributed to electrostatic
ion-acoustic waves (IAWs) coexisting with Alfvénic activity. IAWs are longitudinal electrostatic plasma modes able to propagate along a background magnetic field $\vec{B}_0$, and they have been extensively investigated and reported
theoretically~\cite{Fried1961,Gary1985}, in numerical
simulations~\cite{Valentini2011c}, space environments such as the
solar wind and Earth's
magnetosphere~\cite{Gurnett1977,gary_ionacousticlike_1978,valentini_short-wavelength_2011,Valentini2012,bale_fields_2016},
and in laboratory conditions~\cite{Alexeff1967,Fried1971}. Early spacecraft observations with Helios reported IAWs in the 1--10 kHz range, with electric fields closely aligned with the solar wind magnetic field~\cite{Gurnett1977,Gurnett1979,gurnett_frank_1978}. More recently, a statistical survey based on Solar Orbiter showed that IAWs are the most frequently observed electrostatic mode between 0.5--1 AU, particularly near the perihelion, with mean wave frequencies around 3 kHz and amplitudes near 2.5 mV/m, and that they propagate predominantly along the background magnetic field~\cite{Pisa2021}. Multi-scale studies of solar wind dynamics suggest that kinetic processes can play a role in energy transfer across scales~\cite{Bruno2013,Verscharen2019}. Indeed, recent observations by the Parker Solar Probe have reported signatures of triggered ion-acoustic wave activity, which are strongly associated with core-electron heating~\cite{mozer_core_2022} and proton beams~\cite{verniero_parker_2020} in the near-Sun solar wind. Theoretical analyses estimate that such waves can drive substantial heating of the plasma~\cite{kellogg_heating_2024}, while numerical simulations have suggested that ion-acoustic fluctuations may arise from the nonlinear decay of large-amplitude Alfvén waves~\cite{Araneda2008}, from kinetic instabilities generating non-Maxwellian features~\cite{valentini_short-wavelength_2011}, or from turbulent cascades at kinetic scales~\cite{Vecchio2014}.

The dispersion properties of IAWs are strongly controlled by the electron-to-proton temperature ratio $T_e/T_p$. Linear kinetic theory suggests that the damping of IAWs is very strong when $T_e/T_p\sim1$, but becomes progressively weaker as $T_e/T_p>1$ rises, so they become more easily excited by heat-flux-driven instabilities, allowing IAWs to reach larger amplitudes and to persist in time~\cite{Gary1985,Gary}. Using Ulysses solar wind observations, \citeasnoun{Lin2001} showed that the occurrence of IAW bursts correlates both with the electron heat-flux and with $T_e/T_p$. In particular, IAW activity was found to be more frequent when $T_e/T_p$ was elevated, consistent with the idea that higher electron temperatures reduce Landau damping and allow these waves to develop more easily. Electrons in the solar wind are generally isotropic, and their temperature remains relatively constant across various conditions due to high thermal conductivity~\cite{Newbury1998}. So, variations in $T_e/T_p$ are primarily driven by changes in proton temperature~\cite{Briand2009}. 

Although linear theory predicts that electromagnetic and electrostatic modes are decoupled, in a MbQL description they might exchange energy through the particle distribution, so neglecting IAWs could bias relaxation timescales and heating estimates. Here, we focus on the missing electrostatic contribution, and extend the MbQL formalism by
incorporating a broadband IAW spectrum alongside the customary EMIC
fluctuations and perform self-consistent MbQL simulations to quantify
their combined role in the relaxation of proton temperature
anisotropy. This approach extends previous
studies~\cite{moya_effects_2021,navarro_effects_2022}, addresses a gap
in the literature, as far as we know, of the impact of electrostatic
waves on MbQL evolution, and provides new insight into the multi-scale
energy pathways operating in space and laboratory plasmas.

A key methodological distinction must be kept in mind. MbQL theory
evolves only the macroscopic moments of an assumed velocity
distribution. A diffusive quasilinear theory also
exists~\cite{isenberg_perpendicular_2019}, which solves a phase-space
diffusion equation. The former cannot capture nonlinear phase-space
structures such as particle trapping, resonance plateaus, or
vortex-like distortions, often found in fully kinetic simulations.
Despite this limitation, MbQL offers a computationally economical way
to follow the slow relaxation of temperature anisotropy driven by the
EMIC instability, recasting an otherwise expensive kinetic problem as
a system of ordinary differential equations that can be solved orders
of magnitude faster than fully kinetic models. We note that such
phase-space features can modify quantitative outcomes. For example,
cyclotron-resonant plateaus tend to reduce growth rates and hasten
relaxation toward marginal stability, whereas electron/ion trapping in
IAWs can diminish effective Landau damping and prolong electrostatic
activity, potentially altering relaxation rates and saturation levels.
A detailed analysis of these effects lies beyond the scope of the
present manuscript.

\section{Moment-Based Quasilinear Equations}
We consider a magnetized plasma composed of bi-Maxwellian protons and
isotropic Maxwellian electrons immersed in a background magnetic field
$\vec{B}_0$. Protons are modeled with two different temperatures
perpendicular ($T_{\perp p}$) and parallel ($T_{\parallel p}$) with
respect to $\vec{B}_0$, while electrons are considered an isotropic
thermal bath of constant temperature $T_e$. For waves propagating
along $\vec{B}_0$, the linearized Vlasov-Maxwell equations predict two
independent kind of waves~\cite{krall_trivelpiece,Davidson}: A pure
longitudinal electrostatic wave and a purely transverse wave of
circular polarization whose dispersion relations are, respectively,
\begin{eqnarray}
  0 &= k_\parallel^2 + 2\sum_{\mu=e,p} \frac{\omega_{p\mu}^2}{u_{\parallel\mu}^2} \chi_{\parallel\mu} \,, \label{eq:es-disprel}
  \\
  c^2 k_\parallel^2 &= \omega_k^2 + \sum_{\mu=e,p} \omega_{p\mu}^2\ \chi_{\perp\mu} \,, \label{eq:em-disprel}
\end{eqnarray}
where $c$ is the speed of light, $\omega_k$ is the complex wave frequency, $k$
is its wavenumber along $\vec{B}_0$, and $\chi_{\parallel\mu}$ and
$\chi_{\perp\mu}$ are the susceptibilities of species $\mu$ ($\mu=e$ for
electrons and $\mu=p$ for protons), given by
\begin{eqnarray}
  \chi_{\parallel\mu} &= 1 + \xi_{\mu} Z(\xi_{\mu}) \,,
  \label{eq:chiz}\\
  \chi_{\perp\mu} &= A_\mu + (\xi_{\mu} + A_{\mu} \xi^{-}_{\mu}) Z(\xi^{-}_\mu)\,, \label{eq:chiperp}
\end{eqnarray}
where $\omega_{p\mu}=\sqrt{4\pi e^2 n/m_\mu}$ is the plasma frequency
of species $\mu$, with $e$ the fundamental charge, $n$ the density of
protons and electrons, and $m_\mu$ their masses;
$u_{\parallel\mu} = \sqrt{2k_B T_{\parallel\mu}/m_\mu}$ the parallel
thermal speed, with $k_B$ the Boltzmann constant;
$A_{\mu} = T_{\mu\perp}/T_{\mu\parallel} - 1$ is a measure of the
temperature anisotropy; $\xi_\mu = \omega/k_\parallel u_{\mu\parallel}$ and
$\xi_\mu^{-} = (\omega - \Omega_\mu)/k_\parallel u_{\parallel\mu}$ are
resonance factors, with $\Omega_\mu = q_\mu B_0/m_\mu c$ the
gyrofrequency of species $\mu$, $q_p=-q_e=e$; and $Z$ is the plasma
dispersion function, which is calculated numerically with the Faddeeva
function provided by SciPy. All variables are written in CGS units.

The moment-based quasilinear theory (MbQL) assumes that the
macroscopic parameters evolve adiabatically. In this case, the wave
frequencies $\omega_k = \omega_k(t)$ solve the relevant dispersion
relation, either \eref{eq:es-disprel} or \eref{eq:em-disprel},
instantaneously at all times. By following the procedure described by
e.g. \citeasnoun{Yoon1992} or \citeasnoun{yoon_kinetic_2017}, it can be found that the distribution
function $F_\mu(\vec{v},t)$ evolves adiabatically by following a
diffusion-like equation as
\begin{eqnarray}
  \label{eq:diff}
  \frac{\partial F_{\mu}}{\partial t}
  =
  -\frac{q_{\mu}}{m_{\mu}} \frac{1}{L} \int_{- \infty}^{\infty} dk_\parallel
\left[ \left(1 + \frac{v_{\parallel} k_\parallel}{\omega_{-k}} \right)  \vec{E}_{-k}
- (\vec{v} \cdot  \vec{E}_{-k}) \frac{\vec{k}}{\omega_{- k}} \right] \cdot \frac{\partial
  \delta f_{\mu k}}{\partial \vec{v}}\,,
\end{eqnarray}
where $\vec{E}_k$ is the electric field spectrum with components
perpendicular ($E_{\perp k}$) and parallel ($E_{\parallel k}$) with
respect to $\vec{B}_0$, and $\delta f_{\mu k}$ is a small perturbation
of the distribution function from $F_\mu$ given by
\begin{eqnarray}
\delta f_{\mu k} &= - i \frac{q_{\mu}}{m_{\mu}} \left[ \lambda_{\mu} \frac{e^{-
i \phi}  E_{\perp k}}{\omega_k - v_{\parallel} k_\parallel - \Omega_{\mu}} +
\frac{ E_{\parallel k}}{\omega_k - v_{\parallel} k_\parallel} \frac{\partial
  F_{\mu}}{\partial v_{\parallel}} \right] \,,
\end{eqnarray}
where
\begin{eqnarray}
  \lambda_\mu &= \left(1 - \frac{v_\parallel k_\parallel}{\omega_k} \right) \frac{\partial
F_{\mu}}{\partial v_{\perp}} + \frac{v_{\perp} k_\parallel}{\omega_k} \frac{\partial F_{\mu}}{\partial
v_{\parallel}} \,.
\end{eqnarray}

If $F_\mu(\vec{v})$ is a bi-Maxwellian distribution with thermal speeds $u_{\perp\mu}$ and $u_{\parallel\mu}$, then the MbQL equations can be deduced from
\begin{eqnarray}
\frac{1}{2}\frac{\partial u_{\parallel\mu}^2}{\partial t} &= \int_0^{2\pi} d\phi \int_0^\infty dv_{\perp} v_\perp \int_{-\infty}^\infty dv_{\parallel}  \left[ v_\parallel^2
  \frac{\partial F_{\mu}}{\partial t} \right]\,,
  \\
\frac{\partial u_{\perp\mu}^2}{\partial t} &= \int_0^{2\pi} d\phi \int_0^\infty dv_{\perp} v_\perp \int_{-\infty}^\infty dv_{\parallel}  \left[ v_\perp^2
  \frac{\partial F_{\mu}}{\partial t} \right]\,.
\end{eqnarray}

After some straightforward calculations, the quasilinear evolution of the
perpendicular and parallel thermal speed for protons are then given by
\begin{eqnarray}
  \frac{\partial u^2_{\parallel p}}{\partial t} &= \mathrm{Im} \frac{8}{L}\frac{e^2}{m_p^2} \int_{-\infty}^{\infty}dk_{\parallel} \left[
                                                  \frac{|B_{\perp k}|^2}{c^2k_\parallel^2} \left((\omega_k - \Omega_{p}) \chi_{\perp p} + \omega_k\right) +
                                                  \frac{|E_{\parallel k}|^2}{u_{\parallel p}^2 k_\parallel^2} \chi_{\parallel p} \omega_k
                                                  \right] \,, \label{eq:qlz}
  \\
  \frac{\partial u^2_{\perp p} }{\partial t} &= - \mathrm{Im} \frac{4}{L}\frac{e^2}{m_p^2} \int_{-\infty}^{\infty}dk_{\parallel} \frac{\left|B_{\perp k}\right|^2}{c^2k_\parallel^2} \left[(2i\gamma_k - \Omega_{p})\chi_{\perp p} + \omega_k \right] \,, \label{eq:qlperp}
\end{eqnarray}
where $L$ is the characteristic length of the plasma,
$u_{\perp p}=\sqrt{2k_BT_{\perp p}/m_p}$ is the perpendicular thermal
speed, $|B_{\perp k}(t)|^2 = |B_k(0)|^2 e^{2\gamma_kt}$ is the
spectral transverse magnetic energy consistent with $|B_{\perp k}|^2=|\omega_k/ck_\parallel|^2 |E_{\perp k}|^2$
through the Maxwell equations, and
$|E_{\parallel k}(t)|^2 = |E_k(0)|^2 e^{2\gamma_kt}$ is the spectral
wave energy of longitudinal waves. Here, the wave energies are
assumed to grow exponentially with growth rate
$\gamma_k = \mathrm{Im}(\omega_k)$. The term associated with $|E_{\parallel k}|$ in Eq.~\eref{eq:qlz} is often omitted in other MbQL studies. An important advantage of the MbQL formalism is that it naturally guarantees energy conservation. Here, we assume that electrons have isotropic temperatures $T_{\perp e} = T_{\parallel e} \equiv T_e$, which remain constant through time, so we can focus on the proton dynamics only.

It is important to note that the frequencies accompanying $|E_{\parallel k}|^2$ are taken to satisfy the IAW dispersion relation Eq.~\eref{eq:es-disprel}, whereas those accompanying  $|B_{\perp k}|^2$ satisfy the ACW dispersion relation Eq.~\eref{eq:em-disprel}. Because these branches decouple in the linear approximation and for strictly parallel propagation, their characteristic timescales are different. In particular, the ACW evolves on a proton-cyclotron timescale $\sim\Omega_p^{-1}$, while the IAW evolves on an acoustic timescale $\sim(u_{\parallel p} k_\parallel)^{-1}$. A quasilinear treatment must therefore resolve both the cyclotron and acoustic timescales simultaneously. We return to this point below.

\section{Numerical Results}
Here, we report numerical solutions of the system of differential equations \eref{eq:qlz} and \eref{eq:qlperp}, solved through an
Euler-Cromer scheme. The integrals over wavenumbers are calculated
through a trapezoidal rule, with both integrals calculated in
the range $10^{-3}< v_Ak_{\parallel}/\Omega_p< 12$. The dispersion
relation \Eref{eq:es-disprel} is solved at each time step $t$ using
the \citeasnoun{Muller1956} method to determine the frequencies
$\omega_k$ of IAWs associated to the $E_{\parallel k}$ term, and
\Eref{eq:em-disprel} is solved in a similar way to determine the
frequencies of ACWs associated to the $B_{\perp k}$ term. It is worth
noting that both dispersion relations \Eref{eq:es-disprel} and
\eref{eq:em-disprel} support an infinite number of solutions for
$\omega_k$, and we select only the least damped mode below the proton
gyrofrequency at $k_\parallel\sim0$. From here on, all results are
shown in normalized units, e.g., time is normalized with
$\omega_{pp}^{-1}$, frequencies with $\Omega_p$, wavenumbers with
$\Omega_p/v_A=\omega_{pp}/c$, and the wave spectrum $|E_{\parallel k}|^2$ and
$|B_{\perp k}|^2$ both with respect to
$\Theta_0^2=(Lv_A/4\Omega_p) B_0^2$. For all studied cases, we consider an
uniform background spectrum for both $|B_k(0)|^2$ and
$|E_k(0)|^2$. Although fluctuations arising from turbulence
\cite{matthaeus_turbulence_2021} or inhomogeneities \cite{Azevedo1969}
are known to exist, they are beyond the scope of the present study and
will not be considered here.
\begin{figure}[t!]
    \centering
    \includegraphics[width=1.0\linewidth]{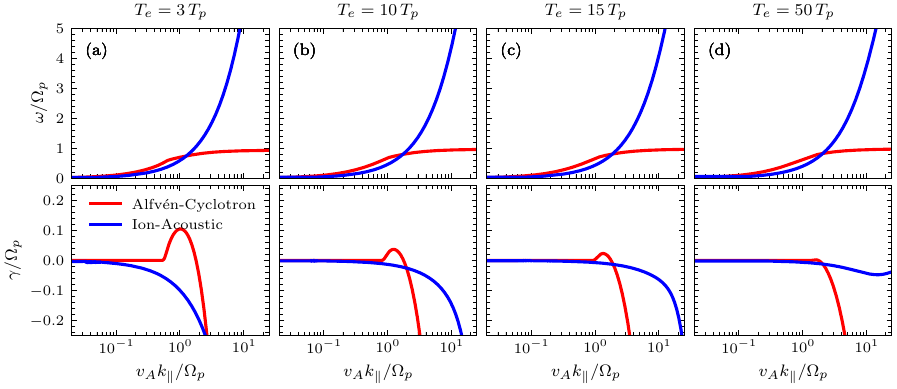}
    \caption{(Top) Frequency and (bottom) growth/damping rate of the Alfvén-Cyclotron (red) and Ion-Acoustic (blue) wave modes as a function of the normalized wavenumber $v_A k_{\parallel}/\Omega_p$, as calculated from the dispersion relation equations \eref{eq:em-disprel} and \eref{eq:es-disprel}. The wavenumbers are shown in logarithmic scale. Each column show results for different temperature ratios $T_e/T_p$, with fixed $T_{p \perp}/T_{p\parallel} = 5$ and $\beta_{e \parallel} = 0.3$. This is equivalent to vary the proton beta as $\beta_{p \parallel} = 0.1$, $0.03$, $0.02$, $0.006$, respectively.}
    \label{fig:disprel-tempratio}
\end{figure}

\Fref{fig:disprel-tempratio} displays the real and imaginary parts of the
complex frequencies of the ACW and IAW modes, both as a function of
the normalized wavenumber $v_A k_\parallel/\Omega_p$ for
$\beta_{\parallel e}=0.3$, and the proton temperature anisotropy as
$T_{\perp p}/T_{\parallel p}=5$, across a range of temperature ratios
$T_e/T_p$. In all cases, both modes have approximately the same
frequency and their damping rates are relatively small for
$v_Ak_{\parallel}/\Omega_p<1$, meaning that a portion of protons may
be able to resonate with both the ACW and IAW modes
simultaneously. Regarding the damping/growth rate $\gamma$, the
spectrum can be broadly divided into three parts, independently of
$T_e/T_p$: one where the system remains relatively stable for
$v_A k/\Omega_p\lesssim 1$, another characterized by resonant
amplification of the ACW at $1\lesssim v_A k/\Omega_p \lesssim 2$, and
a third region where both modes are heavily damped for
$v_A k/\Omega_p \gtrsim 2$. Let us note that the IAW tends to be
marginally stable ($\gamma \approx 0$) until
$v_Ak_{\parallel}/\Omega_p \sim 1$ for all the temperature ratio
considered here. At fixed $\beta_e$, higher values of $T_e/T_p$ is
equivalent to consider higher values of $\beta_{\parallel p}$. Thus,
for $T_e/T_p=3$ (or $\beta_{\parallel p}=0.1$) the ACW mode is excited
more efficiently than for higher values of $T_e/T_p$, corresponding to
an instability typically named as electromagnetic ion-cyclotron (EMIC)
instability~\cite{Shahzad2024}. Also, as $T_e/T_p$ grows, the IAW reduces its damping
rate, being nearly stable at $T_e/T_p=50$, meaning that they may be
more involved in energy transfer in a MbQL simulation.

\begin{figure}[t!]
    \centering
    \includegraphics[width=1.0\linewidth]{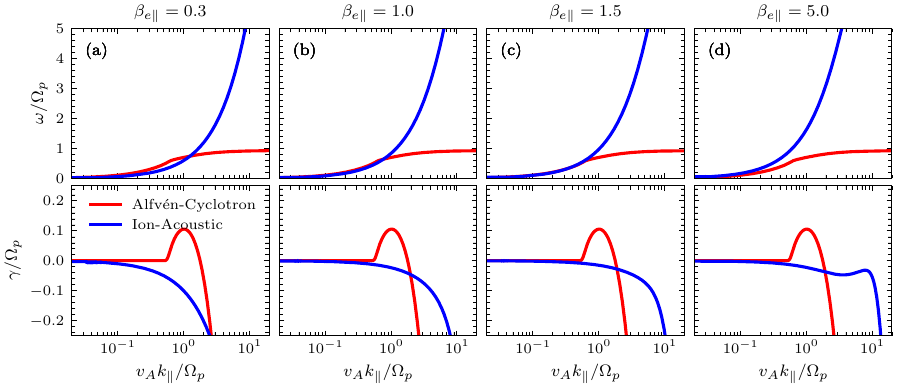}
    \caption{Same as in \Fref{fig:disprel-tempratio}, but for fixed $T_{p \perp}/T_{p\parallel} = 5$ and $\beta_{p \parallel} = 0.1$, and varying $\beta_{\parallel e}$.}
    \label{fig:disprel-betae}
\end{figure}

\Fref{fig:disprel-betae} shows results of the dispersion relation for fixed
$T_{\perp p}/T_{\parallel p}=5$ and $\beta_{\parallel p}=0.1$, and
varying $\beta_{\parallel e}$. Here, we observe that
$\beta_{\parallel e}$ significantly alters the IAW dispersion,
shifting the double-resonance regime towards lower wavenumbers as
$\beta_{e\parallel}$ increases. The damping and growth rate of ACWs do
not change significantly with respect to $\beta_{\parallel e}$, but
IAWs become relatively more stable for higher values of
$\beta_{\parallel e}$ and higher $k_\parallel$. The parameter
conditions used in Figs. \ref{fig:disprel-tempratio} and
\ref{fig:disprel-betae} were selected based on observations from space
plasma environments. In particular, in the inner
heliosphere (e.g., fast-wind streams observed by Parker Solar Probe),
$\beta_{\parallel p}$ commonly falls in the low range
\cite{Halekas2020Electrons,Huang2020Anisotropy,Raouafi2023PSPreview,Kasper2021PRL};
near 1~AU the solar wind typically exhibits
$\beta_{\parallel p} \sim 0.3$ to 1 with median $T_e/T_p \sim 1$ to 3
\cite{Wilson2018SWTemps,Wilson2023Erratum}; and in planetary magnetosheaths
the plasma is frequently high-beta ($\beta_{\parallel p} \gtrsim 1$ to 10)
\cite{Rakhmanova2021MSHReview}. We also note that ion-acoustic waves are
frequently observed between 0.5--1~AU (e.g., Solar Orbiter/RPW/TDS) and have
been linked to heat-flux-rich intervals and localized heating in near-Sun
measurements~\cite{Pisa2021,Graham2021KEW,Mozer2023TIAW}. In the case of the ACW
mode, we adopted a well-established unstable configuration
\cite{moya_study_2011,moya2012,moya_effects_2021} known to be capable
of interacting competitively with the IAW response.

Now we focus on the quasilinear evolution of the plasma due to the
competition of an unstable ACW and the presence of
IAWs. For academic and computational convenience, we
adopt $v_A/c = 0.1$ to reduce simulation time and computational
cost. Notice that this value is unlikely in solar wind and
magnetosphere plasmas, but observed in mildly extreme conditions such
as tenuous strongly-magnetised pre-flare coronal loops in active solar
regions, where constraints from Hinode/XRT give $v_A/c\gtrsim0.1$
\cite{Hudson2008}, and the upstream plasma of low-Mach relativistic
shocks in moderately magnetised jets of AGN or GRB internal shocks,
for which simulations routinely adopt a proper Alfvén speed
$v_A/c\simeq0.1$ \cite{Vainio2003}. For comparison, in typical solar wind conditions one expects $v_A/c \sim 10^{-4}$. In our normalization, the proton cyclotron frequency scales as $\Omega_p/\omega_{pp} = v_A/c$, which implies that the characteristic timescales for ACW–particle interactions in the solar wind would be $\sim 10^{3}$ times longer than those presented here. However, we have found that
the value of $v_A/c$ does not considerably affect the dispersion properties of ACWs. On the other hand, the presence of IAWs introduces
a new temporal scale, and values of $v_A/c$ close to one help to
reduce the number of iteration time-steps needed to achieve a
quasi-stationary state. In the following, the quasilinear simulations
are run with a time-step of
$\omega_{pp}\Delta t=0.01$ for $N=2^{19}$ time-steps (equivalently $\omega_{pp}t_\mathrm{max}\simeq 5240$ or $\Omega_p t_\mathrm{max}\simeq524$.

\begin{figure}[t!]
    \centering
    \includegraphics[width=1.0\linewidth]{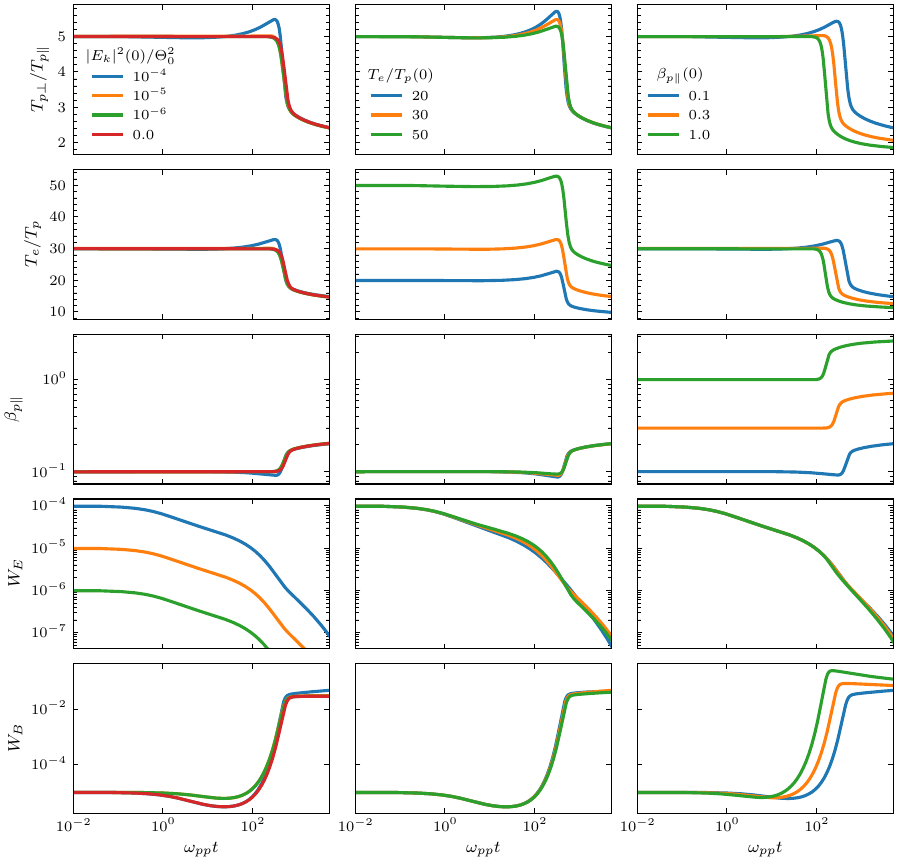}
    \caption{Quasilinear time evolution of the proton temperature
      anisotropy $T_{\perp p}/T_{\parallel p}$, the electron-to-proton
      temperature ratio $T_e/T_p$, the proton beta
      $\beta_{\parallel p}$, and the IAW and ACW total energies
      $W_E$ and $W_B$, respectively. Each column represents a case
      with (left) initial IAW amplitude $|E_k|^2(0) = 10^{-4}$,
      $10^{-5}$, $10^{-6}$, and $0.0$ in units of $\Theta_0^2=(Lv_A/4\Omega_p)B_0^2$; (center) initial electron-to-proton
      temperature ratio $T_e/T_p(0)=20$, $30$, and $50$, and; (right)
      initial proton $\beta_{\parallel p}=0.1$, $0.3$, and $1.0$.  For
      all cases, other initial parameters are
      $T_{\perp p}/T_{\parallel p}(0)=5$, $\beta_{\parallel p}(0)=0.1$,
      $T_e/T_p(0)=30$, $|E_k|^2(0)/\Theta_0^2=10^{-4}$, and $|B_k|^2(0)/\Theta_0^2=10^{-5}$.}
    \label{fig:ql-evolution}
\end{figure}

\Fref{fig:ql-evolution} (left) shows the quasilinear time evolution of
four initial conditions with decreasing IAW amplitude
$|E_k(0)|^2 = 10^{-4}$, $10^{-5}$, $10^{-6}$, and $0.0$. In all cases the initial plasma parameters are the same, with $T_{\perp p}/T_{\parallel p}(0)=5$,
$\beta_{\parallel p}(0)=0.1$, $T_e/T_p(0)=30$, and $|B_k(0)|^2/\Theta_0^2=10^{-5}$.
From top to bottom, each row shows the time evolution of the proton
temperature anisotropy $T_{\perp p}/T_{\parallel p}$, the
electron-to-proton temperature ratio $T_e/T_p$, the parallel proton beta
$\beta_{\parallel p}$, the IAW energy $W_E$, and the ACW energy $W_B$. The wave energies $W_E(t)$ and $W_B(t)$ are obtained by integrating the instantaneous wave amplitudes $|E_{\parallel k}(t)|^2$ and $|B_{\perp k}(t)|^2$, respectively, over all wavenumbers. As shown by Fig.~\ref{fig:disprel-tempratio}, the ACW is unstable in this case. During the quasilinear evolution, this
instability will be reduced by the plasma, which results in a
reduction of $T_{\perp p}/T_{\parallel p}$ over time. Similarly,
$T_e/T_p$ tends to decrease towards thermodynamic equilibrium. During
the process, $\beta_{\parallel p}$ and $|B_k(t)|$ tend to increase, while
$|E_k(t)|$ decreases over time. Although this could be explained by an
energy exchange from the electric field to the protons, notice that
the final $\beta_{\parallel p}$ state is similar in all cases. Also,
previous studies~\cite{yoon_kinetic_2017,moya_effects_2021} have shown
that the increment of $\beta_{\parallel p}$ can occur solely by the
presence of ACWs. Thus, most of the process occurs because of energy
exchange with ACWs, which is seen as the increment of $|B_k|^2$ over
time. Notice, however, that for higher values of $|E_k|^2(0)/\Theta_0^2=10^{-4}$,
there is a slight increment of the temperature anisotropies
$T_{\perp p}/T_{\parallel p}$ and $T_e/T_p$ at $\omega_{pp}
t=100$. This transient increment scales with the initial IAW amplitude and disappears in the special case $|E_k|^2(0)=0$, which is consistent with previously reported quasilinear ACW results~\cite{moya_effects_2021}. Since in the process the IAWs are slightly damped, most of the
electrostatic energy $|E_k|^2$ is absorbed by the plasma, which by
$\omega_{pp} t=100$ has lost two orders of magnitude. Thus, the IAWs
are unable to produce higher values of the temperature anisotropies,
and the process saturates so that the system reduces the instability
as expected at later times. Between times $1\leq \omega_{pp}t\leq 100$,
the magnetic energy also suffers a slight drop and subsequent recovery
of the magnetic field energy, with higher damping than in cases with higher $|E_k|^2(0)$. For lower values of $|E_k|^2(0)$, no significant changes are observed.

\Fref{fig:ql-evolution} (center) presents the quasilinear time
evolution for three values of the electron-to-proton temperature ratio
$T_e/T_p(0)=20$, $30$, and $50$, and for $|E_k(0)|^2/\Theta_0^2=10^{-4}$. Although ACWs can still heat protons in the transverse direction, this process is damped as $T_e/T_p(0)$ increases. However, the electrostatic and transverse magnetic energies undergo only minor changes during the quasilinear evolution.

Similarly, \fref{fig:ql-evolution} (right) shows the quasilinear
evolution for different values of $\beta_{\parallel p}(0)$, retaining
$T_e/T_p(0)=30$ and $|E_k(0)|^2/\Theta_0^2 = 10^{-4}$. Here, the transverse
heating by ACWs is decreased by increasing values of
$\beta_{\parallel p}$. Furthermore, the anisotropic relaxation both in
$T_{\perp p}/T_{\parallel p}$ and $T_e/T_p$ is accelerated by
increasing values of $\beta_{\parallel p}$, while the magnetic field
also gains energy (see $W_B$) faster. The electric field is mostly
unchanged by the value of $\beta_{\parallel p}$. Notice that for
$T_{\perp p}/T_{\parallel p}(0)=5$, an ACW instability is triggered
which is stronger with increasing $\beta_{\parallel p}(0)$. Thus, in
this case, the ACW instability is probably mediating the energy
exchange, which regulates the excess of free energy in the form of
temperature anisotropies.

\begin{figure}[t!]
    \centering
    \includegraphics[width=1.0\linewidth]{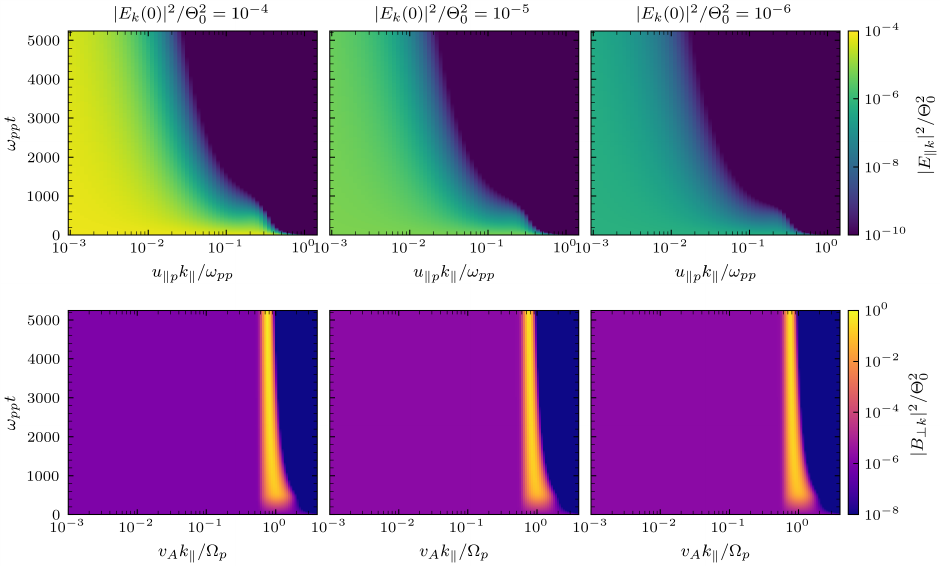}
    \caption{(Top) Time evolution of the electric field $|E_{\parallel k}|^2$ and (bottom)
      magnetic field $|B_{\perp k}|^2$ power spectra for different initial electric
      field amplitudes $|E_k|^2(0)$. Other initial conditions are the
      same as in \Fref{fig:ql-evolution}(left). Both spectra are shown as
      functions of their respective normalized wavenumbers on a
      logarithmic scale, i.e.
      $u_{\parallel p} k_{\parallel} / \omega_{pp}$ for the IAW and
      $v_A k_{\parallel} / \Omega_p$ for the ACW. The colorbar
      indicates the logarithmic intensity of the fields over time in units of $\Theta_0^2$.}
    \label{fig:spectral-fields}
\end{figure}

\Fref{fig:spectral-fields} shows the time evolution of the $|E_k|^2$
and $|B_k|^2$ power spectra, both in units of $\Theta_0^2$, for the same cases as in
\Fref{fig:ql-evolution} (left). Most of the electrostatic energy
$|E_k|^2$ is mostly concentrated at low values of $k_\parallel$, and
the range of wavenumbers where $|E_k|^2$ is higher is reduced over
time, meaning that electrostatic field energy is transferred to the
protons or to the transverse magnetic field at small spatial scales. The electric field overall loses energy over time. In contrast, the magnetic field spectrum $|B_k|$ exhibits amplitude growth around $v_A k_\parallel/\Omega_p=1$, near the wavenumbers where we expect an ACW (or EMIC) instability. Over time, the range of wavenumbers where this amplitude growth occurs is narrowed, but stabilizes around $\omega_{pp} t = 3000$. In this case, energy is transferred from the magnetic field at
$v_A k_\parallel/\Omega_p>1$, and to the magnetic field at
$v_A k_\parallel/\Omega_p\simeq 1$. For lower values of
$v_A k_\parallel/\Omega_p<1$, no changes in the ACWs amplitude are
observed.

\begin{figure}[t!]
    \centering
    \includegraphics[width=1.0\linewidth]{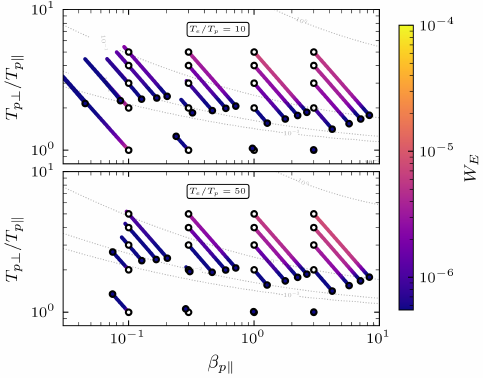}
        \caption{Quasilinear evolution of the proton $\beta_{p\parallel}$ and anisotropy $T_{\perp}/T_{\parallel}$ for two temperatures ratio, $T_e/T_p = 10$ (upper) and $T_e/T_p = 50$ (bottom). In both, initial conditions (white circles) were chosen evenly spaced in the range $0.1\leq \beta_{\parallel}\leq 3$ and $1 \leq T_{\perp}/T_{\parallel} \leq 5$, with an uniform electric wave spectrum power of $|E_k|^2(0)/\Theta_0^2 = 10^{-4}$, an uniform magnetic wave spectrum power of $|B_k|^2(0)/\Theta_0^2 = 10^{-5}$ for all cases. The colorbar represents the instantaneous value of $W_E(t)$. Dashed lines are contours of the proton-cyclotron instability with maximum growth rates $\gamma/\Omega_p = 10^{-3}$ to $10^0$, as calculated from the dispersion relation \eref{eq:em-disprel} considering the corresponding temperature ratio to set $\beta_{e\parallel}$. Colorized circles correspond to the stationary (final) state of the plasma simulations.}\label{fig:beta-aniso}
\end{figure}

\Fref{fig:beta-aniso} shows the time evolution of
$\beta_{p\parallel}$, $T_{p\perp}/T_{p\parallel}$, and $W_E$ for an
ensemble of initial simulations for electron-to-proton temperature
ratios $T_e/T_p = 10$ and $T_e/T_p = 50$. A set of numerical simulations with evenly log-spaced initial conditions were chosen in the range of $0.1\leq \beta_{\parallel}\leq 3$ and $1 \leq T_{\perp}/T_{\parallel} \leq 5$, marked with white circles, with a uniform wave spectrum of power $|E_k|^2(0)/\Theta_0^2 = 10^{-4}$ and $|B_k|^2(0)/\Theta_0^2 = 10^{-5}$ for all cases. The colored lines represent the quasilinear $\beta_{\parallel p}$--$T_{\perp p}/T_{\parallel p}$ path of the system, and the color represents the instantaneous electric field energy $W_E$. The colored circles represent the final state of the system at $\omega_{pp}t = 10^{3}$ when the system has reached a quasi-stationary state. Moreover, contours of maximum growth rate $\gamma_{max}/\Omega_p$ of the ACW instability are included as
segmented lines, where states below these lines correspond to stable
plasma conditions according to the numerical solutions of the
dispersion relation \Eref{eq:em-disprel}, so their states are not
expected to change over time. However, notice that simulations
starting below the instability threshold, e.g., at
$\beta_{p\parallel}(0) = 0.1$ and $T_{\perp}/T_{\parallel} = 1$, the anisotropy can grow up to very high values above the instability
thresholds in the case of $T_e/T_p=10$. For higher values of
$\beta_{\parallel p}$ or for $T_e/T_p=50$, this anisotropy growth
still occurs, but is saturated rapidly and before exciting an
instability. In any case, the total electrostatic energy $W_E$
decreases from $10^{-4}$ to $\approx 10^{-6}$ at the end of the
simulations.

Several quasilinear studies have shown that even initially isotropic, nominally stable proton populations can undergo strong perpendicular heating and evolve into unstable regimes when exposed to an ambient magnetic fluctuation spectrum~\cite{isenberg_perpendicular_2019,moya_effects_2021}. In the case of \fref{fig:beta-aniso} (upper), while both ACWs and IAWs contribute to the wave spectrum, the observed perpendicular heating of initially isotropic protons beyond instability thresholds at $\beta_{\parallel p}\sim 0.1$ is most consistent with quasilinear interactions with ACWs, which can efficiently energize protons through cyclotron resonance. However, the efficiency of this heating decreases as either $T_e/T_p$ or $\beta_{\parallel p}$ increase. This trend is consistent with theoretical expectations, since at high electron-to-proton temperature ratios, IAWs experience weaker ion damping and thus couple more strongly to electrons, while the proton cyclotron resonance with ACWs becomes less effective at sustaining perpendicular heating beyond the instability threshold.

On the other hand, initial plasma configurations above the instability
thresholds, e.g., $\beta_{p\parallel}(0) \geq 1.0$ and
$T_{\perp}/T_{\parallel} = 5$ for both cases, the instability is
rapidly damped, and the plasma reaches a quasi-stationary state where
$10^{-3}\leq \gamma/\Omega_p\leq 10^{-1}$. In these cases, the plasma
responds less efficiently to the electric perturbation, and the EMIC instability becomes predominant. As a result, the system does not reach significant levels of anisotropy, and
the fluctuation is rapidly damped.

\section{Discussion and Conclusions}
In this work, we have employed the moment-based quasilinear theory to investigate the combined effects and interplay of ion-acoustic waves (IAWs) and Alfvén-Cyclotron waves (ACWs) in regulating the proton temperature anisotropy in a plasma of bi-Maxwellian protons and Maxwellian
electrons. While linear kinetic theory predicts that these two wave branches are decoupled under parallel propagation with respect to a background magnetic field, in the quasilinear regime both interact with the particle distributions through energy exchange. Unlike similar studies
\cite{moya_effects_2021,navarro_effects_2022}, our model fully retains the kinetic behavior of both plasma species, avoiding the cold electron approximation, to coherently study the role of IAWs. Electrons are treated as a thermal reservoir, enabling us to focus on the quasilinear proton heating channels.

In general, plasmas with thermally isotropic protons $T_{\perp p}/T_{\parallel p}=1$ are incapable of exciting ACW instabilities under the linear regime. However, under the quasilinear approach, a sufficient amplitude of ACWs can provide enough free energy to increase $T_{\perp p}/T_{\parallel p}$, especially at low $\beta_{\parallel p}$. This heating can drive initially stable distributions into unstable regimes, consistent with earlier quasilinear studies~\cite{moya_effects_2021}. The efficiency of this process diminishes as either $\beta_{\parallel p}$ or the electron-to-proton temperature ratio $T_e/T_p$ increase. As the electron populations becomes hotter relative to protons, IAWs experience weaker ion damping and absorb a larger fraction of the fluctuation energy through Landau resonance, thereby reducing the share available to ACW-driven proton heating. Once ACW instabilities are triggered, the system evolves toward a quasi-stationary state where growth rates relax to $\gamma/\Omega_p <10^{-1}$. For sufficiently large $\beta_{\parallel p}$, $T_e/T_p$, or instability growth rates, ACWs become inefficient drivers of perpendicular proton heating, leaving IAWs as the dominant dissipation channel and underscoring their active role in regulating proton heating. More broadly, wave-mediated energy transfer has also been highlighted in laboratory contexts such as laser-plasma interactions~\cite{Wani2019,Kamboj2021}, illustrating the universality of this mechanism across plasma regimes.

Our results also show that IAWs can transiently accelerate the production of proton temperature anisotropy when their initial amplitude is sufficiently large. This short-lived effect reflects the rapid absorption of electrostatic energy as the IAWs are damped, after which the ACWs regain dominance and drive the long-term evolution toward reduced anisotropy and quasi-stationary saturation. A qualitatively similar sequence has been reported in other plasma contexts where an initial two-stream electrostatic instability saturates rapidly and is subsequently overtaken by a slower, but more robust, electromagnetic instability~\citeasnoun{Lopez2020}. In our case, although the physical mechanisms differ, the parallel is clear as the IAWs may govern the very early stage of the dynamics, but the sustained regulation of anisotropy is ultimately controlled by ACWs.

We recognize that the use of a moment-based quasilinear approach is
limited, since this model assumes that the proton distribution function evolves by maintaining its bi-Maxwellian shape, thus discarding processes like phase-space diffusion and the formation of BGK waves. In this framework, any resonant contribution is expressed in terms of the previously introduced dispersion relations, since we do not have direct access to the distribution function itself, but only to the slow evolution of its macroscopic parameters. Nonetheless, our approach also provides a framework to study separately the effects of different waves, such as IAWs and ACWs combined, which is difficult to separate in a more general model, fully non-linear numerical simulations, or spacecraft data analysis. 

In the astrophysical context of collisionless shocks, electrostatic activity in the form of either IAWs or electron-acoustic waves, may play a role in the development of electrostatic shocks acceleration, driven by the electron thermal pressure gradient~\cite{Marcowith2016}. While this effect is not expected to extend to large-scale environments such as the intergalactic medium, it is relevant at the Debye scale, where electrostatic shocks have been reported in the transition region of the Earth's collisionless bow shock, along with associated BGK-type structures~\cite{Bale2002}. 


We note that the assumption of uniform initial spectra for $|E_k(0)|^2$ and $|B_k(0)|^2$ is a simplification. While a power-law spectrum would better reflect turbulent conditions and modify the transfer rates, the qualitative conclusion that ACWs drive perpendicular heating while IAWs regulate it should remain robust. In future work, we aim to extend this framework to include more realistic features, such as a hotter second electron population, electron-proton relative drift allowing for current-driven instabilities \cite{Buneman1959,Yoon2010}, or the presence of heavier ion species, such as $\alpha$-particles \cite{Araneda2009,Araneda2011} or oxygen ions, for which observations have revealed highly anisotropic distributions~\cite{Reisenfeld2001,Wilhelm2007}. Moreover, a comprehensive validation of the MbQL framework against both fully kinetic simulations and in-situ spacecraft observations represents a natural next step beyond the scope of the present study, and we intend to pursue it in future investigations.

\ack
We thank the support of ANID Chile through FONDECyT grants No. 1240281 (P.S.M.) and No. 1240697 (R.E.N.). M.A.Q. acknowledges D. M. Rivera and M.V. Coello for their insightful discussions, which contributed to the development of this work. This research was supported by the International Space Science Institute (ISSI) in Bern, through the ISSI International Team project 24-612: Excitation and Dissipation of Kinetic-Scale Fluctuations in Space Plasmas. We would also like to thank the Local Organizing Committee of the 17th Latin American Workshop on Plasma Physics LAWPP-2025, held in Santiago, Chile, for a great and fruitful scientific meeting, in which an early version of this work was presented. The authors declare no competing financial or non-financial interests.

\section*{Data-Availability Statement}
All data supporting the findings are generated by numerical simulations based on the equations described in this manuscript. The complete input files and raw outputs are available from the corresponding author upon reasonable request. This study involves no human participants, human data or tissue, and no animal subjects; therefore, ethics approval and informed-consent requirements do not apply.

\section*{References}
\bibliographystyle{jphysicsB-nouniquename} 
\bibliography{references}

\end{document}